\documentclass[pra,twocolumn,a4paper,showpacs,floatfix]{revtex4} % double column
\usepackage{graphicx} \newcommand{\figspace}{\vspace*{2ex}}
\usepackage{amsmath}
\begin{document}
\title{Electronic transport in a Cantor stub waveguide network}
\author{Sheelan Sengupta, Arunava Chakrabarti} \affiliation{Department of Physics, University of Kalyani, Kalyani, West Bengal 741 235, India}
\author{S.Chattopadhyay}\affiliation{Department of Physics, Hoogly Mohsin College, Chinsurah, West Bengal 712 101, India}
\begin{abstract}
We investigate theoretically, the character of electronic eigenstates and transmission properties of a one dimensional array of stubs with Cantor geometry. Within the framework of real space re-normalization group (RSRG) and transfer matrix methods we analyze the resonant transmission and extended wave-functions in a Cantor array of stubs, which lack translational order. Apart from resonant states with high transmittance we unravel a whole family of wave-functions supported by such an array clamped between two-infinite ordered leads, which have an extended character in the RSRG scheme, but, for such states the transmission coefficient across the lead-sample-lead structure decays following a power-law as the system grows in size. This feature is explained from renormalization group ideas and may lead to the possibility of trapping of electronic, optical or acoustic waves in such hierarchical geometries.
\end{abstract}
\pacs{73.21.La,73.63.Kv,85.35.Be}
\maketitle
%----------------------------------------------------------------------------
\section{Introduction}
%---------------------------------------------------------------------------
Electron transmission in quantum wires with various geometric structures has been an interesting area of research in mesoscopic physics in recent times \cite{pd00,fs89,jbx92,psd94,jrs96,jov98,vp02,jmc02,pao03}. Apart from presenting an interesting problem in itself, this area of research has profound implications in applied physics and materials science, particularly, in designing nano-devices with exotic electronic properties. 
\vskip .2in
\noindent
Nano-technology, as its present advanced stage, makes it possible to fabricate a variety of quantum structures such as quantum dots, wires and rings. The surface of solids can be designed with appreciable control to serve as ideal templates for the formation of low-dimensional nano-structures with desired geometry \cite{lwb97}. Nano-devices such as one-dimensional wires can be experimentally prepared by controlling self-organized defects at the surface \cite{ns91}. Apart from the path-breaking experiments, from the point of view of a theoretician, an interesting observation is that, often the basic principle underlying the operation of these fabricated nano-systems can be modeled in a quite simple manner, using quantum mechanics and elementary solid state physics. One such example is the study of electron transport in model quantum wires with dangling side branches (stubs) \cite{fs89,jbx92,psd94,jrs96,jov98,vp02,jmc02,pao03} of finite length. Such `dangling' stubs can be grafted along a nano-wire with any specified ordering. The presence of stubs in a quantum wire has been shown to lead to transistor action by Sols et al \cite{fs89}. The length of the stubs can be controlled by applying a gate voltage, and a change in these lengths result in a non-trivial change in the spectrum of the system \cite{fs89,jbx92}. The spectrum and band structure of an array of stubs were studied by Deo and Jayannavar \cite{psd94} as well. They pointed out that, unlike a system of potential scatterers, a system of geometrical scatterers (such as an array of stubs) exhibit complete band formation in the conductance with only a few such stubs. Several studies on the quantum waveguide transport using model stubbed nano-wires then followed. Shi and Gu \cite{jrs96} examined the electronic conductance of a quantum waveguide with side-branches taking into account the impedance factors for geometric and potential scatterers. Vasseur et al \cite{jov98} studied simple tight binding models of one dimensional comb structures of simple metals. In a similar spirit Pouthier and Girardet \cite{vp02} have investigated the conductance of  a monatomic nano-wire containing a random distribution of side-grafted nano-clusters of atoms. Cervero and Rodriguez \cite{jmc02} modeled a quantum wire by a random array of delta potentials and extended their work to investigate the effect of positionally correlated potentials on the spectral behaviour. Orellana et al  \cite{pao03} stuck to the tight binging model to investigate the electronic transport through a quantum wire with a side quantum dot array. Quite recently, Cao et al \cite{yc03} have studied the electronic transport in a comb structure with a mesoscopic ring threaded by a magnetic flux. While the Schr{\"o}dinger equation approach \cite{jbx92,psd94,jrs96,jmc02,pao03,yc03} is a natural choice for the calculation of the conductance, discrete difference equation approach \cite{jov98,vp02,pao03} also finds its justification by the recent advancement in atom manipulation technology in which one can place individual atoms on the substrate by using a scanning tunnel microscope tip as tweezers \cite{jov98}.
\vskip .2in
\noindent
Though the transmission characteristics of a stubbed quantum wire are quite sensitive to the geometry of the structure, little effort has been given to study the spectral characteristics when the arrangement of the stubs deviates from being periodic. One interesting area, where in-depth studies in this regard are really lacking is, when the stubs are arranged in a hierarchical pattern. In this communication we undertake a detailed study of the electronic states and transport in a model one-dimensional quantum wire with stubs of finite length grafted on it following a triadic Cantor sequence (a one-dimensional fractal) \cite{fc92}. Apart from providing direct experimental evidence for localized and self-similar fracton mode displacements \cite{fc92}, interesting theoretical results indicate the presence of multiple fracton and phonon regimes in a triadic Cantor sequence \cite{ap94}. Also, the Cantor distribution of dielectric layers has recently found application in the design of guided-wave optical or microwave filters \cite{ass00}. Comparatively speaking, the electronic properties of a Cantor quantum waveguide network is unattended so far, though recent tight binding results on the electronic properties of a Cantor lattice with point scatterers indicate exotic features \cite{ss04}, being markedly different in several aspects from the conventional quasi-periodic lattices of the Fibonacci class \cite{mk83}. In the case of a stubbed waveguide network the possibility of tuning the stub-lengths experimentally may provide an opportunity for practically `seeing' some of the peculiar features already encountered in the theoretical studies of such Cantor-like substitutional sequences \cite{ss04,bl97}.
\vskip .2in
\noindent
With this motivation we undertake a detailed analysis of electronic transmission and the nature of wave-function in a Cantor stub waveguide (CSW) network using the real space renormalization group (RSRG) and the transfer matrix (TM) method. We discretize the Schr{\"o}dinger equation to map the original sequence of stubs onto an equivalent one-dimensional chain consisting of point-like scatterers placed at the nodes where a stub meets the backbone. This facilitates the application of RSRG method. Using RSRG with TM method we unravel a family of extended (resonant) eigenstates even through the underlying lattice lacks translational order. It is not unusual that, by controlling the length of a stub appropriately one can have an infinite number of wave-vectors of the propagating electron which fall in the spectral gaps of an infinite CSW. However, a very interesting feature of a CSW array is that, the same wave-vectors correspond to perfectly allowed (in some cases, even periodic) wave-functions for a CSW of arbitrarily large, but finite length clamped between semi-infinite ordered leads. The most remarkable character of such states is that, the nearest neighbour hopping integral on the mapped lattice [Fig.1(b)] exhibit a two-cycle fixed point under  RSRG operation, indicating an `extended' character (in the RSRG sense), and yet the transmission coefficient of the `lead-sample-lead' system decays following a power law, as the sample grows in size. We term such states as `atypically extended'. This result is in total contrast to recent studies on a Fibonacci quantum waveguide network \cite{gjj99,sc04}.
\vskip .2in
\noindent
In what follows we describe our work. In section II, we introduce the model and the method, section III consists of a discussion of the overall transmission characteristics with an emphasis on the resonant eigenstates and in section IV we talk about the `atypical' eigenfunctions and the possibility of the trapping an excitation in such a quantum waveguide network. In section V we draw conclusion. 
%----------------------------------------------------------------------------
\section{The model and the method}
%-----------------------------------------------------------------------------
\begin{figure}
  \centering \figspace
  \centerline{\includegraphics[width=0.95\columnwidth]{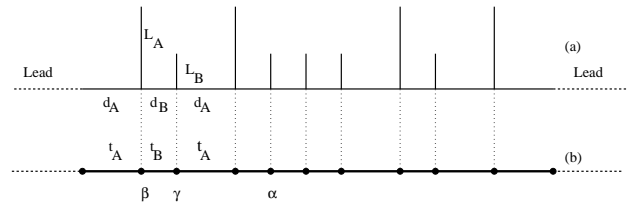}}
\caption{\label{fig1}
(a) Second generation Cantor array of quantum stub clamped in between two semi-infinite leads. (b) The equivalent one-dimensional chain.} 
\end{figure}
%----------------------------------------------------------------------------
A triadic Cantor sequence is a binary arrangement of two elements, $A$ and $B$ for example \cite{ap94}. The sequence grows in a recursive manner, starting with a seed $A$, using a substitution rule, $A \rightarrow ABA$ and $B \rightarrow BBB$. The first few generations are, $G_0 = A$, $G_1 = ABA$, $G_2 = ABABBBABA$, and so on. It is to be noted that such a sequence differs from conventional binary quasi-periodic sequence \cite{mk83} in the sense that, in the thermodynamic limit the sub-clusters of $B$ will span the entire one-dimensional space with the elements $A$ acting as impurities located at specific sites determined by the sequence. 
\vskip .2in
\noindent
We model our system by attaching a series of stubs perpendicularly to a one dimensional wire. Stubs of lengths $L_A$ and $L_B$ are arranged in a Cantor sequence with inter-stub spacings $d_A$ and $d_B$ as shown in Fig.1(a). We consider an electron entering the system from one side of the nano-wire with wave-vector $q$. The wave-function in the $n$th segment (chosen to be the segment between the $n$th and the $(n+1)$th stubs along the $X$-direction) is given by, 
\begin{equation}
\psi_n(x) = A_n e^{iq(x-x_n)} + B_n e^{-iq(x-x_n)}
\end{equation}
\noindent
The wave-function in the $n$th stub (along the $Y$-direction) is written as, 
\begin{equation}
\phi_n(y) = C_n e^{iq(y-L_n)} + D_n e^{-iq(y-L_n)}
\end{equation}
\noindent
The above equations are written using a local coordinate system in the spirit of Ref. \cite{gjj99}. It is then simple to match the wave-functions and their derivatives at the nodes to arrive at the matrix equation:
\begin{eqnarray}
\left(\begin{array}{c}
A_{n+1} \\ B_{n+1}\end{array}\right)
& = & 
Q_{n+1,n} \left(\begin{array}{c}
A_{n} \\ B_{n}
\end{array}\right)
\end{eqnarray}
\noindent
where,
\begin{eqnarray}
Q_{n+1,n} & = & \left(\begin{array}{cc}
1-\frac{i\cot qL_n}{2} & -\frac{i\cot qL_n}{2} \\ \frac{i\cot qL_n}{2} & 1+\frac{i\cot qL_n}{2}
\end{array} \right) \times \nonumber \\
& &
\left(\begin{array}{cc}e^{iqd_n} & 0 \\ 0 & e^{-iqd_n} 
\end{array} \right)
\end{eqnarray}
\noindent
Instead of dealing with the matrices $Q$ we discretize Eq.(3) into a set of difference equation much in the spirit of Poincare mapping [J. B. Sokoloff, in Ref. \cite{mk83}]. The resulting set of equations assume a rather simple form,
\begin{equation}
(E - \epsilon_n)\Psi_n = t_{n,n+1}\Psi_{n+1} + t_{n,n-1}\Psi_{n-1}
\end{equation}
\noindent
where, $\Psi_n$ is the amplitude of the wave function at the node $n$. This allows us to deal with the problem in the same manner as we do for a tight binding electron of energy $E$ moving in a one dimensional lattice with on-site potential $\epsilon_n$ for the $n$th site and the nearest neighbour hopping integrals 
$t_{n,n\pm 1}$. In our case the quantity $(E-\epsilon_n)$ assumes three different values depending on the nearest neighbour environment. The values are,
\begin{eqnarray}
E - \epsilon_\alpha & = & 2\cot qd_B + \cot qL_B \nonumber \\
E - \epsilon_\beta & = & \cot qd_A + \cot qd_B+ \cot qL_A \nonumber \\
E - \epsilon_\gamma & = & \cot qd_A + \cot qd_B+ \cot qL_B 
\end{eqnarray}
\noindent
The nearest neighbour hopping integrals on the mapped lattice are given by $t_A=1/\sin qd_A$ and $t_B=1/\sin qd_B$ across the lengths $d_A$ and $d_B$ respectively. Eq. (15) is easily cast into the matrix form, viz.
\begin{displaymath}
\left(\begin{array}{c}
\Psi_{n+1} \\ \Psi_n\end{array}\right)=M_n \left(\begin{array}{c}
\Psi_n \\ \Psi_{n-1} 
\end{array}\right)
\end{displaymath}
\noindent
where we define the three $2 \times 2$ transfer matrices $M_n$ as, 
\begin{eqnarray}
M_\alpha & = & \left(\begin{array}{cc}
\frac{(E-\epsilon_\alpha)}{t_B} & -1\\ 1 & 0 
\end{array} \right) \nonumber \\
M_\beta & = & \left(\begin{array}{cc}
\frac{(E-\epsilon_\beta)}{t_B} & -\frac{t_A}{t_B} \\ 1 & 0 
\end{array} \right) \nonumber \\
M_\gamma & = & \left(\begin{array}{cc}
\frac{(E-\epsilon_\gamma)}{t_A} & -\frac{t_B}{t_A} \\ 1 & 0 
\end{array} \right)
\end{eqnarray}
\noindent
according to the nature of the $n$th site (i.e. $\alpha$, $\beta$, $\gamma$). With the parameters $\epsilon_\alpha$, $\epsilon_\beta$, $\epsilon_\gamma$ and $t_A$, $t_B$ and the transfer matrices $M_\alpha$, $M_\beta$, and $M_\gamma$ we are now in a position to analyze the electron states and transmission properties of the equivalent one-dimensional lattice [Fig.1(b)] using the RSRG method.
%-----------------------------------------------------------------------------
\begin{figure}
\centering \figspace
\centerline{\includegraphics[width=0.95\columnwidth]{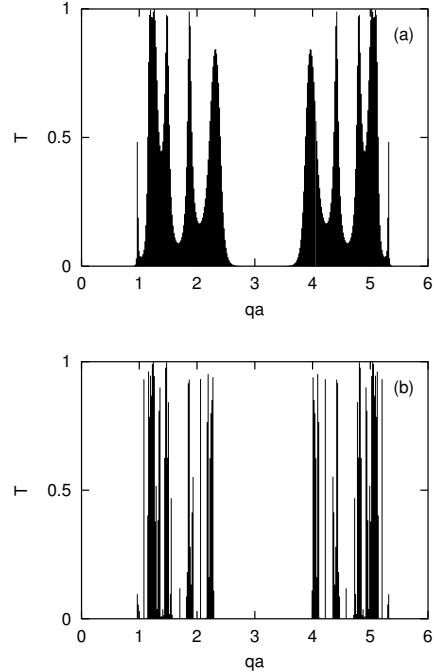}}
\caption{\label{fig2}
Variation of $T$ with $q$ when $d_A=d_B=a=1$, $L_A=0.5$, $L_B=1$.} 
\end{figure}
%----------------------------------------------------------------------------
\section{Transmission coefficient and the resonant eigenstates}
%-----------------------------------------------------------------------------
We now present the results of our calculation in relation to the transmission properties of finite CSW of arbitrary lengths. Let us, without losing any physics, stick to a particular model of the waveguide array, consisting of equispaced stubs of two different lengths $L_A$ and $L_B$. The equivalent one dimensional lattice then consists of two different on-site potentials $\epsilon_A$ and $\epsilon_B$, and a constant hopping integral $t$. This geometry is easily achieved by setting $\epsilon_\alpha=\epsilon_\gamma=\epsilon_B$, $\epsilon_\beta=\epsilon_A$ and $d_A=d_B=a$ so that, $t_A=t_B=t$. The transfer matrices are now of two types, viz. $M_\alpha=M_\gamma=M_A$ and $M_\beta=M_B$. The product transfer matrix for a CSW at the $l$th generation is given by $M_l=M_{l-1}M_B^{3^{l-1}}M_{l-1}$, with $M_1=M_AM_BM_A$.
\vskip .2in
\noindent
In order to evaluate the transmission coefficient of a Cantor stub waveguide (CSW) at any generation, we place the equivalent one-dimensional sample between two semi-infinite leads. The lead, in a tight binding description, is characterized by identical on-site potentials $\epsilon_0$ and constant hopping integral $t_0$. The transmission coefficient for such a system then can be worked out in terms of the traces and anti-traces \cite{md90} of the relevant transfer matrices and for a $l$th generation Cantor stub array it is given by,
\begin{equation}
T_l=\frac{4\sin ^2 qa}{(z_l\cos qa-y_l)^2+x_l^2\sin ^2 qa}
\end{equation} 
\noindent
where, $x_l=TrM_l$, $y_l=M_l(2,1)-M_l(1,2)$ and $z_l=M_l(1,1)-M_l(2,2)$ respectively. The trace and anti-traces can be evaluated recursively by using the equations,
\begin{eqnarray}
x_{l+1} & = & U_{F_l-1}(x_B)[x_lw_l-x_B]-U_{F_l-2}(x_B)[x_l^2-2] \nonumber \\
y_{l+1} & = & U_{F_l -1}(x_B)[y_lw_l+y_B]-U_{F_l -2}(x_B)x_ly_l \nonumber \\
z_{l+1} & = & U_{F_l -1}(x_B)[z_lw_l+z_B]-U_{F_l -2}(x_B)x_ly_l \nonumber \\
w_{l+1} & = & U_{F_l -1}(x_B)[w_l^2-2]-U_{F_l -2}(x_B)[x_lw_l-x_B]\nonumber\\
\end{eqnarray}
\noindent
Here, $F_l=3^{l-1}$ and $U_m$ is the mth order Chebyshev polynomial of the second kind. $x_B=Tr(M_B)$, $w_l=Tr(M_BM_l)$, $y_B=M_B(2,1)-M_B(1,2)$, and $z_B=M_B(1,1)-M_B(2,2)$ respectively. The initial values of the traces and anti-traces are
\begin{eqnarray}
x_1 &=& \frac{(E-\epsilon_A)^2(E-\epsilon_B)}{t^3}-\frac{2(E-\epsilon_A)}{t}-\frac{(E-\epsilon_B)}{t}\nonumber \\
y_1 &=& \frac{2[(E-\epsilon_A)(E-\epsilon_B)]-1]}{t^2}\nonumber \\
z_1 &=& \frac{(E-\epsilon_A)^2(E-\epsilon_B)}{t^3}-\frac{2(E-\epsilon_A)}{t}+\frac{(E-\epsilon_B)}{t}\nonumber \\
w_1 &=& \frac{(E-\epsilon_A)^2(E-\epsilon_B)^2}{t^4} \nonumber \\
& & -\frac{4(E-\epsilon_A)(E-\epsilon_B)}{t^2}+2 \nonumber
\end{eqnarray}
\noindent
In Fig.2 we present the variation of the transmission coefficient against the wave vector ranging from zero to $2\pi/a$ for the second and fourth generation Cantor stub waveguide system. It is seen that there is a wide central gap is the spectrum with the finite transmission zones clustering around $qa=\frac{\pi}{2}$ and $\frac{3\pi}{2}$. With increasing size, the two sub-bands around $\frac{\pi}{2}$ and $\frac{3\pi}{2}$ start getting fragmented [Fig.2(b)], giving rise to multiple closely spaced zones of high transmission separated from each other by gaps. Keeping all other parameters fixed as in Fig.2, a change in the value of one of the stub lengths, say, $L_A$, is found to produce notable changes in the transmission spectrum as well as in the nature of the eigenfunction. This aspect may be used in a possible experimental setup in order to control the transport at a specified value of the wave-vector. Let us clarify by giving an example. We set $U_2(x_B)=0$. The corresponding value of the wave-vector satisfies the equation
\begin{displaymath}
\cos qa=\pm \frac{1}{3}
\end{displaymath}
\noindent
It can easily be checked that the above choice of $q$ (or $qa$) makes $M_B^3=\pm I$. This implies that an arbitrarily large Cantor sequence of stubs of $L_A$ and $L_B$ becomes indistinguishable from a periodic array of the triplet $L_AL_BL_A$ If $q=\frac{1}{a}\cos ^{-1}(\pm \frac{1}{3})$ happens to be in the allowed energy spectrum of an infinite periodic chain of the triplet, we get an extended eigenfunction, even though the original Cantor sequence lacks translational invariance. In Fig.3 we plot the variation of $T$ with $L_A$ for $q=\cos ^{-1}(-\frac{1}{3})$ and $a=1$, $L_B=1$ within half a period. An interesting feature in Fig.3 is that for such a value of the wave-vector to correspond to a finite appreciable value of the transmission coefficient, $L_A$ has to exceed a threshold value. We have not been able to determine the threshold exactly, but careful numerical search has revealed that, as one deals with longer and longer generation segments, the threshold remains close to $q=0.2$ (in the unit of $a$). For bigger generations change in the threshold value of $L_A$ becomes practically un-noticible
%-------------------------------------------------------------------------------
\begin{figure}
\centering \figspace
\centerline{\includegraphics[width=0.95\columnwidth]{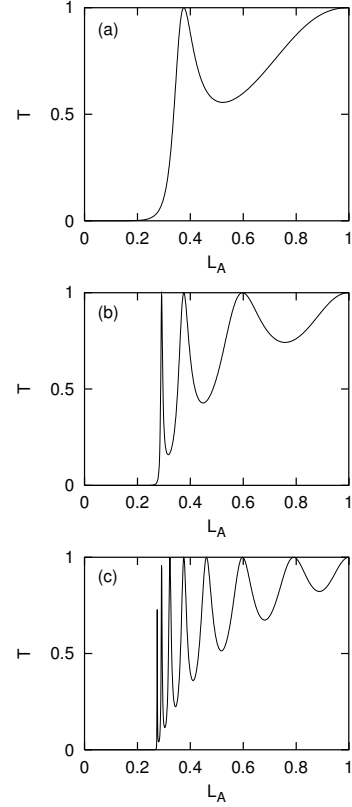}}
\caption{\label{fig3}
$T$ vs $L_A$ with $q=\cos ^{-1}(-\frac{1}{3})$, $d_A=d_B=a=1$, and $L_B=1$ (a) second generation, (b) third generation (c) fourth generation.} 
\end{figure}
%----------------------------------------------------------------------------
\vskip .2in
\noindent 
It is instructive to look at the distribution of amplitudes of the wave function for the above value of $q$ for a semi-infinite stub array. It is well known that the clusters of impurities in a one dimensional chain are likely to give rise to unscattered (resonant) eigenstates. This is true for a random dimer model \cite{dhd90} and its extensions, and for a certain class of quasi-periodic chains as well \cite{ac94}. In the same spirit one can search for $q$-values corresponding to extended wave-function in a CSW network by making higher powers of $M_B$ equal to identity matrix. The method is well known \cite{dhd90, ac94}, and we do not repeat it to save space. In the example discussed above, the amplitude profile reflects a perfectly extended character, and assumes an interesting lattice-like distribution \cite{ss04} for a specific value of $L_A=0.5$. In Fig.4 we show such a lattice-like profile of an extended eigenstate. Other combinations of the wave-vector and the system parameters also give rise to similar observations. However whether the amplitude distribution resemble the lattice or not depends on the specific values of the parameters, the wave-function may be of an extended nature.
%-------------------------------------------------------------------------------
\begin{figure}
\centering \figspace
\centerline{\includegraphics[width=0.95\columnwidth]{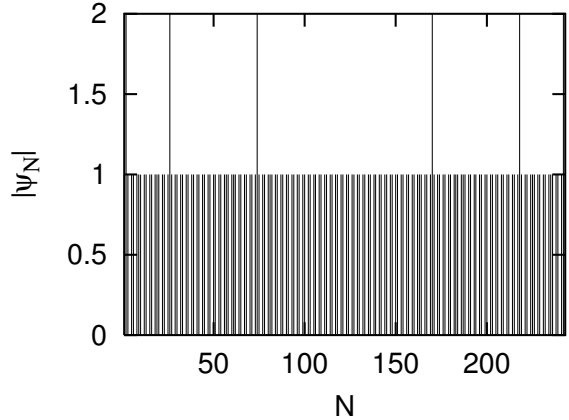}}
\caption{\label{fig4}
Lattice-like profile of an extended eigenstate where $q=\cos ^{-1}(\frac{1}{3})$. We have taken $d_A=d_B=a=1$, $L_A=0.5$ and $L_B=1$.}
\end{figure}
%----------------------------------------------------------------------------
\section{Atypical ``extended states'' and anomalous transport}
%-------------------------------------------------------------------------------
We begin this section by noting that an infinite Cantor stub array can be renormalized into a scaled version of itself simply by using the growth rule in the opposite direction. In terms of the equivalent one-dimensional chain this implies the decimation of a subset of sites. The resulting RSRG recursion relations for the on-site potentials and the hopping integrals are found to be,
\begin{eqnarray}
\epsilon^\prime_\alpha & = & \epsilon_\alpha+ \frac{2t_B^2(E-\epsilon_\alpha)}{w} \nonumber \\
\epsilon^\prime_\beta & = & \epsilon_\beta+ \frac{t_A^2(E-\epsilon_\beta)}{v}+  \frac{t_B^2(E-\epsilon_\alpha)}{w} \nonumber \\
\epsilon^\prime_\gamma & = & \epsilon_\gamma+ \frac{t_A^2(E-\epsilon_\gamma)}{v}+ \frac{t_B^2(E-\epsilon_\alpha)}{w} \nonumber \\
t^\prime_A & = & \frac{t_A^2t_B}{v} \nonumber \\
t^\prime_B & = & \frac{t_B^3}{w}
\end{eqnarray}
\noindent  
where, $v=(E-\epsilon_\beta)(E-\epsilon_\gamma)-t_B^2$ and $w=(E-\epsilon_\alpha)^2-t_B^2$. It is well known that a flow of the hopping integrals under successive RSRG steps provide information about the nature of the wave-function. For example, for an extended wave-function the hopping integrals never flow to zero under any number of RSRG iterations whereas, for a localized state at least one of the two hopping integral will flow to zero under RSRG. However, for the latter case one has to ensure that the chosen energy does not fall into a gap in the spectrum of the infinite lattice.
\vskip .2in
\noindent
We stick to the discussion of a model consisting, as before, of two different equispaced stubs of lengths $L_A$ and $L_B$. An RSRG treatment of this model needs the general recursion relations (10) described before. The present model can be retrieved from the general model described before by setting $d_A=d_B=a$ and $\epsilon_\alpha=\epsilon_\gamma \ne \epsilon_\beta$. We now have $t_A=t_B=\frac{1}{\sin qa}$. Let us now choose $E-\epsilon_\alpha=0$ which results in the equation
\begin{equation}
\cot qL_B+2\cot qa=0
\end{equation}
\noindent
This condition is easily be achieved, for a given value of the lattice spacing $a$, by appropriately choosing the stub length $L_B$. For example, with $a=1$ and $qa=(2n+1)\pi/2$, $n=0,1,2,...$, $L_B$ can assume values equal to $\frac{1}{2n+1}$ (in unit of $a$) for which the above equation is satisfied. The length of the other stub $L_A$, of course remains free to be chosen. Even though we start up with the special model, on decimation, we land up with a mixed one and one has to use the set of recursion relation (10) for drawing conclusion. Iterating the set of Eq.(10) we observe that $\epsilon_\alpha$ and $\epsilon_\gamma$ remain fixed at their initial values while $t_A(=t_B)$ exhibit a two cycle fixed point behaviour, viz,
\begin{displaymath}
t_A^{''}=-t_A^{'}=t_A
\end{displaymath}
\begin{displaymath}
t_B^{''}=-t_B^{'}=t_B
\end{displaymath}
\noindent
Most interestingly $\epsilon_\beta$ increases on successive renormalization following the prescription 
\begin{displaymath}
\epsilon_\beta(n)=2\epsilon_\beta(n-1)-3\epsilon_\alpha,
\end{displaymath}
\noindent
$n$ indicating the stage of renormalization. This implies that an electron propagating in the Cantor stub network with above specification will experience, in a renormalization group sense, progressively higher potential barriers resulting in a decay in the value of the transmission coefficient as the system grows in size.
%-------------------------------------------------------------------------
\begin{figure}
\centering \figspace
\centerline{\includegraphics[width=0.95\columnwidth]{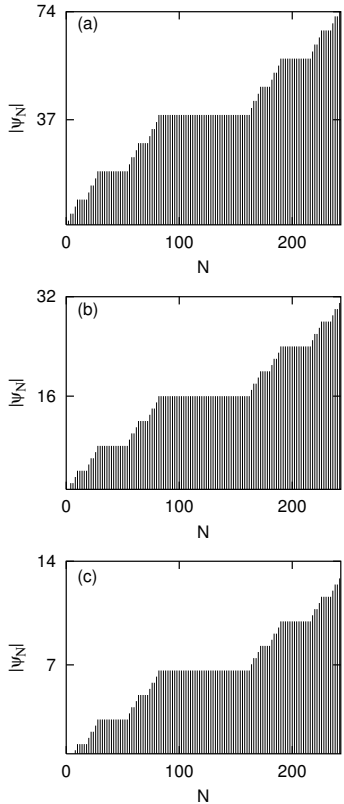}}
\caption{\label{fig5}
Amplitude $\Psi_N$ growth pattern for (a) $L_A=0.25$, (b) $L_A=0.5$ and (c)
 $L_A=.75$ for $q=\frac{\pi}{2}$ (in unit of inter-lattice spacing).
 Here, $L_A=d_A=d_B=1$.}
\end{figure}
%-----------------------------------------------------------------------------
A pertinent question in this regard is, whether $q=(2n+1)\pi/2$ belongs to the spectrum of an infinite or semi-infinite chain. The answer is `no'. With respect to the amplitude on the first site, the amplitudes at distant sites keep on increasing for larger and larger system and consistent solutions of the difference equation (5) is not possible to obtain in this case. However, it is interesting to note, and simple to check that, the same set of all the parameters yields a perfectly allowed, even a periodic distribution of $\psi_i$ when any arbitrary generation of a Cantor segment is clamped between two semi-infinite perfect leads, where the leads are described by identical scatterers placed periodically. The growth of the potential $\epsilon_\beta$ under RSRG implies that an electron released in the middle of a large cluster of $B$ nodes will experience high potential walls on either side and will be ultimately trapped in this well if the system size is large enough. This fact co-exists with the fixed point behaviour of the nearest neighbour hopping integrals which usually characterize an extended eigenstate. We thus come across a set of, what we may call, `atypical' eigenstates.
\vskip .2in
\noindent
Both the growth of the amplitude and the decaying transmission coefficient can easily be estimated if we note that for $q=(2n+1)\pi/2$, and $d_A=d_B=L_B=1$, we get $M_B^2=-I$, so that any arbitrarily large generation of a cantor segment of $A$'s and $B$'s become identical, from the standpoint of the propagating electron, to a periodic array of the unit cells `$AB$'. the transfer matrix across a finite $l$th generation array now reads
\begin{displaymath}    
P_l=\left(\begin{array}{cc}
(-1)^l(N_{AB}(l)+1) & (-1)^{l+1}\\ (-1)^{l-1} & 0
\end{array}\right)
\end{displaymath}
\noindent
where, $N_{AB}(l)$ is the number of $AB$-blocks left in the $l$th generation after the central clusters of $B$ yielding identity matrices. This leads to the power law decay in the transmission coefficient as a function of $N_{AB}(l)$, viz,
\begin{displaymath}
T_l\sim \frac{1}{N_{AB}^2(l)}
\end{displaymath}
%-----------------------------------------------------------------------
\begin{figure}
\centering \figspace
\centerline{\includegraphics[width=0.75\columnwidth,angle=0]{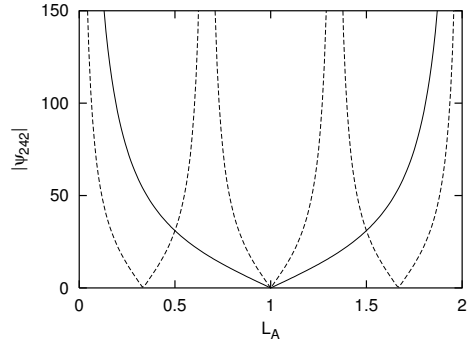}}
\caption{\label{fig6}
Variation of the maximum value of the amplitude in a Cantor waveguide as a
 function of the stub length $L_A$. The solid and the dashed curves correspond
 to (a) $q=\frac{\pi}{2}$ and (b) $q=\frac{3\pi}{2}$. Other parameters are 
$L_B=1$, $d_A=d_B=1$. We show $\mid\Psi_{242}\mid$ up to 150 for clarity.}
\end{figure}
%----------------------------------------------------------------------------
\noindent
Before ending this section, it is worth mentioning that, by tuning the length of the other stub $L_A$ one can reduce the growth of the wave-functions for $q$-values which fall in the spectral gap of a semi-infinite Cantor sequence of stubs. The amplitude of such a wave-function usually exhibits a slowly increasing pattern interspaced with flat constant regions corresponding to the periodic clusters of $B$'s in between the $A$'s. The maximum amplitude for $q=\frac{\pi}{2}$, say, at any generation $l$ occurs at the $(3^l-1)$th site as can easily be checked. In Fig.5 we exhibit the flattening of the amplitude growth pattern for three different values of $L_A$. By changing $L_A$ continuously from 0 to 1, we get a smooth decay of the largest value of the amplitude occurring in any generation until with $L_A=L_B=1$, we get a completely periodic distribution of $\psi_i$. At this stage the lattice becomes indistinguishable from a periodic one. Fig. 6 shows how the largest value of the amplitude in a $243$-stub array drops from large values to $\pm 1$. 
%----------------------------------------------------------------------------
\begin{figure}
\centering \figspace
\centerline{\includegraphics[width=0.75\columnwidth,angle=0]{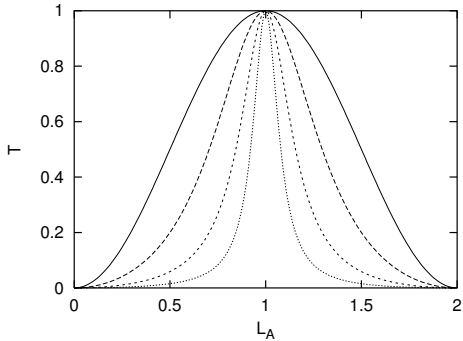}}
\caption{\label{fig7}
Variation of transmission coefficient as a function of $L_A$ for $q=\frac{\pi}{2}$. We have taken $L_B=1$, $d_A=d_B=1$. Solid, dashed, broken and dotted curve correspond to first, second, third and forth generation CSW respectively.}
\end{figure}
%------------------------------------------------------------------------------
\vskip .2in
\noindent
The same aspect is reflected in the variation of $T$ as a function of the stub-length ($L_A$, for example). Fig. 7 shows that perfect resonance occurs at $L_A=1$ only which represents a periodic array of stubs. For all other values of $L_A$ the transmittance decays as the system grows in size (i.e. as $N_{AB}$ increases).  The `switch-over' to complete transparency occurs sharply at $L_A=1$ only when the system is infinitely large i.e. an `off' to `on' state can be obtained as $L_A$ is varied across the value unity.
%-----------------------------------------------------------------------------
\section{Conclusion}
%-----------------------------------------------------------------------------
In conclusion, we have used the renormalization group idea and a transfer matrix method, to examine the electronic states and transport in finite and infinite stubbed waveguide network constructed following a triadic Cantor geometry. Apart from usual extended eigenfunctions which correspond to complete transmission, such a waveguide network when clamped between two perfectly ordered semi-infinite leads, is found to support `atypical' states which may display a periodic pattern of amplitude and yet the end-to-end transmission decays as the system grows in size. We provide our explanation from a renormalization group point of view.
%------------------------------------------------------------------------------
 
\end{document}